\documentclass[sigconf,natbib=true,anonymous=false,screen]{acmart}

\usepackage{algorithm}
\usepackage{algorithmic}
\usepackage{hyperref}
\usepackage{graphicx}
\usepackage{subfig}
\usepackage{subcaption}
\usepackage{amsmath,amsfonts}
\usepackage{array}
\usepackage{diagbox}
\usepackage{booktabs} 
\usepackage{multirow}
\usepackage{makecell}
\usepackage{tablefootnote}
\usepackage{arydshln}
\usepackage{tikz}
\usepackage{longtable}
\usepackage{makecell}
\usepackage{xcolor}
\newcommand{\myparagraph}[1]{\noindent \textbf{#1:}}

\DeclareMathOperator*{\argmax}{arg\,max}
\usepackage{newfloat}
\usepackage{listings}

\AtBeginDocument{%
  }

\copyrightyear{2025}
\acmYear{2025}
\setcopyright{cc}
\setcctype{by}
\acmConference[SIGIR '25]{Proceedings of the 48th International ACM SIGIR Conference on Research and Development in Information Retrieval}{July 13--18, 2025}{Padua, Italy}
\acmBooktitle{Proceedings of the 48th International ACM SIGIR Conference on Research and Development in Information Retrieval (SIGIR '25), July 13--18, 2025, Padua, Italy}\acmDOI{10.1145/3726302.3730110}
\acmISBN{979-8-4007-1592-1/2025/07}

\definecolor{azure}{rgb}{0.0, 0.5, 1.0}

\begin{document}

\title{Unsupervised Corpus Poisoning Attacks in Continuous Space for Dense Retrieval}


\author{Yongkang Li}
\email{y.li7@uva.nl}
\orcid{0000-0001-6837-6184}
\affiliation{%
  \institution{University of Amsterdam}
  \city{Amsterdam}
  \country{The Netherlands}
}

\author{Panagiotis Eustratiadis}
\orcid{0000-0002-9407-1293}
\email{p.efstratiadis@uva.nl}
\affiliation{%
  \institution{University of Amsterdam}
  \city{Amsterdam}
  \country{The Netherlands}
}

\author{Simon Lupart}
\orcid{0009-0008-2383-4557}
\email{s.c.lupart@uva.nl}
\affiliation{%
  \institution{University of Amsterdam}
  \city{Amsterdam}
  \country{The Netherlands}
}

\author{Evangelos Kanoulas}
\orcid{0000-0002-8312-0694}
\email{e.kanoulas@uva.nl}
\affiliation{%
  \institution{University of Amsterdam}
  \city{Amsterdam}
  \country{The Netherlands}
}


\begin{abstract}
This paper concerns corpus poisoning attacks in dense information retrieval, where an adversary attempts to compromise the ranking performance of a search algorithm by injecting a small number of maliciously generated documents into the corpus. Our work addresses two limitations in the current literature. First, attacks that perform adversarial gradient-based word substitution search do so in the discrete lexical space, while retrieval itself happens in the continuous embedding space. We thus propose an optimization method that operates in the embedding space directly. Specifically, we train a perturbation model with the objective of maintaining the geometric distance between the original and adversarial document embeddings, while also maximizing the token-level dissimilarity between the original and adversarial documents. Second, it is common for related work to have a strong assumption that the adversary has prior knowledge about the queries. In this paper, we focus on a more challenging variant of the problem where the adversary assumes no prior knowledge about the query distribution (hence, unsupervised). Our core contribution is an adversarial corpus attack that is fast and effective. We present comprehensive experimental results on both in- and out-of-domain datasets,
focusing on two related tasks: a top-1 attack and a corpus poisoning attack. We consider attacks under both a white-box and a black-box setting. Notably, our method can generate successful adversarial examples in under two minutes per target document; four times faster compared to the fastest gradient-based word substitution methods in the literature with the same hardware.
Furthermore, our adversarial generation method generates text that is more likely to occur under the distribution of natural text (low perplexity), and is therefore more difficult to detect.\looseness=-1
\end{abstract}

\ccsdesc[500]{Information systems~Retrieval models and ranking}
\ccsdesc[500]{Computing methodologies~Natural language processing}

\keywords{Dense Retrieval, Adversarial Attack, Corpus Poisoning }

\maketitle

\section{Introduction}

Dense retrieval~\cite{KarpukhinOMLWEC20_DPR} has become a widely used paradigm in information retrieval (IR), utilizing neural language models to encode queries and documents~\cite{sigir/AzzopardiCKMTRAACREHHKK24}. However, such neural models are susceptible to adversarial attacks \citep{Szegedy13Intriguing}, making adversarial robustness research an important topic in IR. Adversarial attacks in IR typically aim to compromise the ranking performance of a retrieval model (e.g.,~\cite{10.1145/3576923_PRADA}). In this paper, we specifically focus on corpus poisoning attacks (e.g.,~\cite{ZhongHWC23_Poisoning,li2025reproducinghotflip,su2024corpus}), where an adversary attacks by injecting maliciously generated documents into a corpus. The attacker aims to promote uninformative documents and maximize their visibility in the top-ranked results of arbitrary search rankings. We assume that existing documents in the corpus have already been encoded and indexed, and therefore we do not have edit access i.e., we may not \textit{replace} one document with another. Instead, we can only generate new documents to add to the corpus, analogous to how search engines like Google and Bing continuously index newly added web documents, making them available for retrieval.

Contemporary poisoning methods aim to pollute the corpus with documents that not only achieve high rankings, but also are nonsensical to users. It is worth noting that under this threat model \textit{imperceptibility is neither required nor feasible}~\cite{chen-etal-2022-adversarial}. The attack is considered successful when a user encounters the adversarial document positioned at the top, reads it, and perceives it as useless.
Most previous studies have focused on gradient-based word substitution, e.g., models based on HotFlip~\cite{EbrahimiRLD18_hotflip,song2020adversarial,ZhongHWC23_Poisoning,su2024corpus,10.1145/3576923_PRADA}. Such methods first duplicate an existing document in the corpus, and then iteratively replace individual tokens with new ones, adversarially generated to maximize the retriever's error.
However, this not only results in a significant time complexity~(\cite{abs-2402-12784_Vec2Text_Understanding} report 2 hours of search time for 50 tokens using an NVIDIA A100 GPU),
but also induces a misalignment of objectives, as each replacement of a single token occurs in the lexical space, while retrieval itself computes the representation of the entire document in the embedding space.
Bridging this gap is challenging, as there are discontinuities in the adversarial generation process, e.g., decoding token embedding samples from the language distribution of the decoder~\cite{0003XS23_Sentence}.

\begin{figure}[t]
\centering
  \includegraphics[width=\columnwidth]{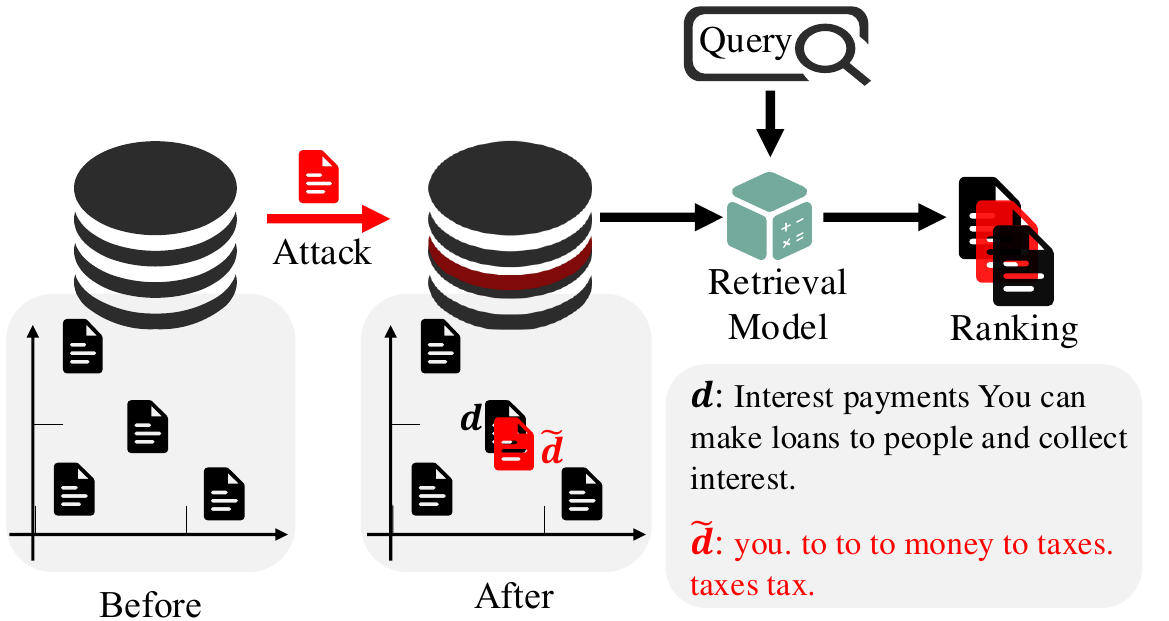}
  \caption{Illustration of our unsupervised corpus poisoning attack under our threat model. We attack a retriever's ranking performance by generating uninformative documents with high relevance scores. For example, encoding the original document \(d\) and its adversarial counterpart \(\tilde{d}\) with SimLM produces similar embeddings, but \(\tilde{d}\) is nonsensical.
}
  \label{fig:task illustration}
\end{figure}

Moreover, most attacks in dense retrieval commonly make the assumption of a target query at the time of attack, that is used to inform how documents are corrupted. To name a few, AGGD\cite{su2024corpus}, IDEM\cite{ChenHY0S23_IDEM}, PRADA~\cite{10.1145/3576923_PRADA}, PAT~\cite{10.1145/3548606.3560683_Order-Disorder}, MCARA~\cite{10.1145/3583780.3614793_MCARA}, TARA~\cite{Liu0GR0FC23_TARA}, as well as the aforementioned corpus poisoning attacks.
We claim that this is a strong assumption from a practical standpoint, as we cannot always rely on knowing the target queries in advance, as well as from a scientific standpoint, as it raises concerns of overfitting the attack models on specific queries.
In this paper, we introduce a more realistic and challenging scenario, termed \textit{unsupervised} corpus poisoning (Figure~\ref{fig:task illustration}). In this context, ``unsupervised'' refers to the fact that there is no prior knowledge about the query distribution at the time of attack, and the attack method itself is only informed by the target document.

We address these two current limitations and propose a corpus poisoning attack that operates directly in the embedding, rather than lexical space.
Our method consists of two main components: a reconstruction model and a perturbation model. The reconstruction model can recover a document from its contextual token-level embeddings. The perturbation model is trained to
maintain the geometric distance between the original and adversarial document embeddings, while also maximizing the token-level dissimilarity between the original and adversarial documents.
We examine two types of adversarial attacks that differ in their criteria for target document selection: one that corrupts the top-1 document of an arbitrary ranking (the query that produced the ranking is unknown), and one that corrupts the $k$ most ``central'' documents in a corpus, with the hypothesis that these documents affect a lot of queries.

Our method is a direct improvement over state-of-the-art~(SOTA) HotFlip-based methods. It performs up to par in white-box attack scenarios, but demonstrates stronger transferability properties in a black-box setting. Our attack is both fast and effective, generating successful adversarial examples at four times the speed of the fastest HotFlip-based approach~\cite{li2025reproducinghotflip}. Moreover, the incorporation of a reconstruction model ensures that our generated outputs closely mimic natural documents, resulting in significantly lower perplexity compared to HotFlip-based methods, making them more difficult to detect by perplexity-based filtering.

Finally, we briefly discuss that the computational efficiency of our method enables the possibility of adversarial training, by using the generated adversarial documents as negative samples. We do not present extensive experimentation in this direction, but enough to suggest to the reader that it is a promising direction for future work.\looseness=-1

\section{Related Work}

\myparagraph{Dense Retrieval}
Following the initial success of dense retrieval models with DPR~\cite{KarpukhinOMLWEC20_DPR}, 
recent advancements  include topic-aware sampling (TAS-B~\cite{HofstatterLYLH21_tasb_dense_retrieval}), unsupervised training with intermediate pseudo queries (Contriever~\cite{izacard2021contriever}),  data augmentation under diversity constraints (DRAGON+~\cite{LinALOLMY023_dragon}), and pre-training with representation bottleneck (SimLM~\cite{WangYHJYJMW23_SimLM}). Our work investigates the vulnerability and robustness of these SOTA models.

\myparagraph{Word Substitution Attacks}
Our method belongs to the family of word substitution ranking attacks~\cite{Wu_2022}, similar to PRADA~\cite{10.1145/3576923_PRADA}, MCARA~\cite{10.1145/3583780.3614793_MCARA} or TARA~\cite{Liu0GR0FC23_TARA}, as well as the family of corpus poisoning attacks such as Order-Disorder~\citep{10.1145/3548606.3560683_Order-Disorder} and HotFlip~\citep{EbrahimiRLD18_hotflip,ZhongHWC23_Poisoning,su2024corpus,li2025reproducinghotflip}. However, these methods assume prior knowledge about a query distribution that guides training, and perturbation search. Our work introduces the novel, \textit{unsupervised} setting, and is uniquely positioned separately from these methods, as we make no assumptions about what kind of queries correspond to the attacked rankings/corpus. The only training signal for our attack method is the document itself. To contextualize our work within the existing literature, we mention that our work is in a similar direction as~\citet{ZhongHWC23_Poisoning}, except (i) no queries are used during training, (ii) perturbation search happens in the embedding space, and (iii) our approach is significantly faster and generates adversarial documents with lower perplexity.\looseness=-1

\myparagraph{Embedding Space Perturbations}
Our work makes use of a reconstruction model, for which we draw inspiration from Vec2Text-based approaches~\citep{MorrisKSR23_vec2text,abs-2402-12784_Vec2Text_Understanding}. We build upon the Vec2Text paradigm by learning instance-wise optimal perturbations, rather than being restricted to additive random noise. Furthermore, our motivation to work in the embedding space, and not on a token level, stems from recent work that leverages Transformer models to generate more powerful attacks, e.g., BERT-ATTACK~\cite{li-etal-2020-bert-attack}. 

\myparagraph{Imperceptibility of Attacks}
There is a line of adversarial IR work that corrupts documents with word substitution  attacks~\cite{10.1145/3576923_PRADA,10.1145/3583780.3614793_MCARA,Liu0GRFC24_multi_granular,Liu0GR0FC23_TARA}, aiming to maintain imperceptibility. 
This imperceptibility refers to preserving the original document's semantics while boosting its ranking during attacks. However, in agreement with~\cite{chen-etal-2022-adversarial}, we argue that imperceptibility is neither necessary nor practical from the user's perspective. 
This is because an attack succeeds only when the user actively reads the top-ranked adversarial document and perceives it as nonsensical,
We found this assumption to be compatible with the ``realistic attack'' angle of our work.

\section{Methodology}

In this section, we first introduce the foundational concepts of dense bi-encoder models. Then, we describe our unsupervised corpus poisoning adversarial attack settings, which lead to two conditions for effective adversarial documents. Based on these conditions, we design an optimization process for generating adversarial content, utilizing both a reconstruction model and a perturbation model.

\subsection{Preliminaries}
The task of dense retrieval concerns scoring a collection of documents, i.e., corpus, $D =\{d_1,d_2,...,d_{|D|}\}$ according to their relevance against a query, $q\in{Q}$. To do so, queries and documents are projected as vectors onto an embedding space by a neural language encoder, and relevance is defined as the dot product, cosine similarity, or L2 Euclidean distance of these vectors.
We denote $E(\cdot)$ as the encoder, and its output is token-level embeddings for the entire document, \(e_d= E(d) \in\mathbb{R}^{\left|d\right|\times\hbar} \), where \(\hbar\) is the hidden dimension of the retrieval encoder, typically \(\hbar=768\).  \(d\) is any document in \(D\), and  \(\left|d\right|\) denotes the number of tokens in $d$,  where \(d =\left\{t_1,t_2,...,t_{\left|d\right|}\right\} \) after tokenization. Each  \(t_i\) is a token from the vocabulary \(V\), and the size of \(V\) is \(\left|V\right|\).

Note that we use the same encoder for queries and documents (i.e., weight-sharing), though different encoders could be used. Since we focus on a query-independent formulation, this design hyperparameter is not crucial.
For scoring \(\text{sim}(\cdot)\) during retrieval, we rely solely on the $\texttt{[CLS]}$ token as the document embedding, while the specific scoring function depends on the retriever.
However, for reconstruction, we utilize all token-level embeddings.

\myparagraph{Problem Formulation}\label{sec:Problem Formulation}
For unsupervised corpus poisoning attacks,
we base ourselves on a white box attack setting where we know the document encoder~\(E(\cdot)\) and similarity function \(\text{sim}(\cdot)\). We inject adversarially generated documents into the corpus, $D$, that satisfy two necessary conditions:

\begin{itemize}
    \item \textbf{Embedding Similarity Condition}:  The adversarial document~\(\tilde{d}\) should be as similar as possible to the target document~\(d_i\) in the embedding space, ensuring a high ranking.
    \item \textbf{Semantic Irrelevance Condition}: The adversarial document~\(\tilde{d}\) should be irrelevant to the target document~\(d_i\) from a human perspective.
\end{itemize}
A successful attack occurs only when the user actively reads the top-ranked adversarial document~\(\tilde{d}\)
and perceives it as uninformative.

\subsection{Reconstruction Model}\label{sec:RM}
\begin{figure}[tbp]
\centering
  \includegraphics[width=\columnwidth]{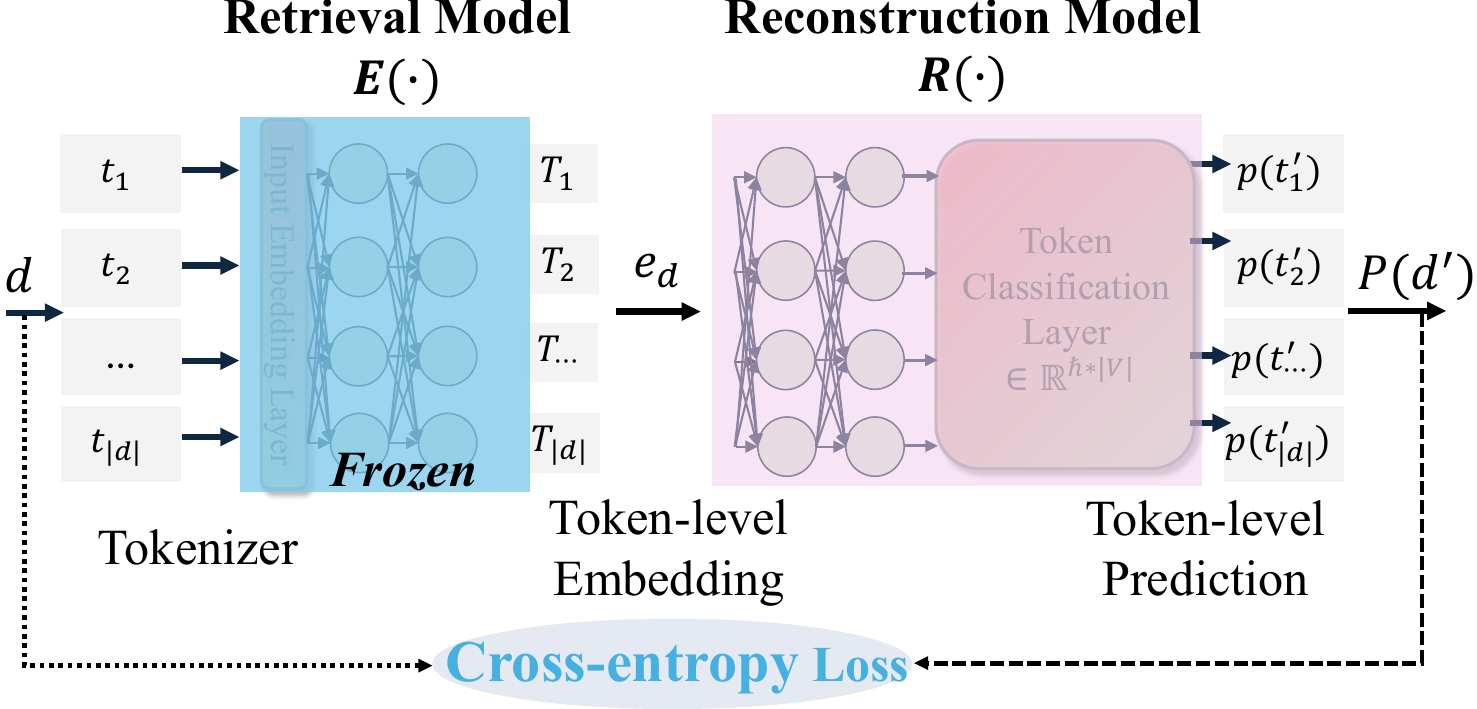}
  \caption{The  training pipeline for the reconstruction model.}
  \label{fig:Reconstruction Model}
\end{figure}

Adversarial attacks in Natural Language Processing are challenging due to the discrete nature of words, where optimizing a single token with word substitution by its gradients is not consistent with optimizing the whole document. In contrast, the continuous nature of the embedding space allows for gradient-based optimization on the whole document. To bridge this gap, we thus develop a reconstruction model enabling us to perturb embeddings directly -- where gradients of documents can be computed -- rather than manipulating discrete tokens.
More specifically, we train a reconstruction model that is able to recover original tokens from retrieval contextual token-level embeddings, such that \(d\simeq  R\left( E \left(d\right)\right)\), with \(e_{d} = E (d) \in \mathbb{R}^{\left|d\right|\times\hbar} \).
Figure~\ref{fig:Reconstruction Model} shows the details of training, where forward propagation formula is as follows:
\begin{align}
\begin{split}
    P\left( d^{\,\prime} \mid e_{d} ; R  \right) = R\left( e_{d} \right) 
         =  R\left( E\left(d\right) \right) 
\end{split}
\end{align}
where  \( P\left( d^{\,\prime} \mid e_{d} ; R  \right) \in \mathbb{R}^{\left|d\right|\times\left|V\right|}\) represents the token-level predicted probabilities within the token space \(V\), abbreviated as \( P\left( d^{\,\prime}  \right) \). We can get each predicted token-ids of reconstructed document \(d^{\,\prime}\) through an \(\argmax(\cdot)\) function: \(d^{\,\prime} = \argmax_{p(t^{\,\prime}_i) \forall t^{\,\prime}_i \in V } P\left( d^{\,\prime} \right)\).

In our design, the reconstruction model~\(R(\cdot)\) consists of a multi-layer transformer encoder structure combined with a multi-class classification layer as the last layer.
The number of classes in the last layer should correspond to the number of tokens in \(E(\cdot) \), which is 30,522 for BERT-based retrieval models. During the training, we freeze the retrieval model \(E(\cdot) \) and train \(R(\cdot) \) by \textbf{minimizing} a cross-entropy loss as \underline{r}econstruction \underline{m}odel~(\textbf{RM}) loss:
\begin{align}
    L_{RM} = - \sum^{\left|D\right|}{ d\cdot \log\left( P\left( d^{\,\prime} \mid e_{d} ; R  \right)\right)} 
\end{align}

After training, we can encode a document at the token level using the retriever and then use the reconstruction model to recover the text from its contextual embeddings, which means \(d^{\,\prime} \simeq  d\). 

\subsection{Adversarial Generation Optimization}

As we mentioned above in the attack problem formulation part, a qualified adversarial document~\(\tilde{d}\) needs to satisfy both the embedding similarity condition and the semantic irrelevance condition.
For the embedding similarity condition, given a target document, we can optimize the Euclidean distance of embeddings between the adversarial document~\(\tilde{d}\) and the target document~\(d\). 
And the semantic irrelevance condition can be achieved by optimizing the number of common tokens between the adversarial document and the target documents. The fewer common tokens the two documents share, the more different their semantic will be.

\begin{figure}[tbp]
\centering
  \includegraphics[width=\columnwidth]{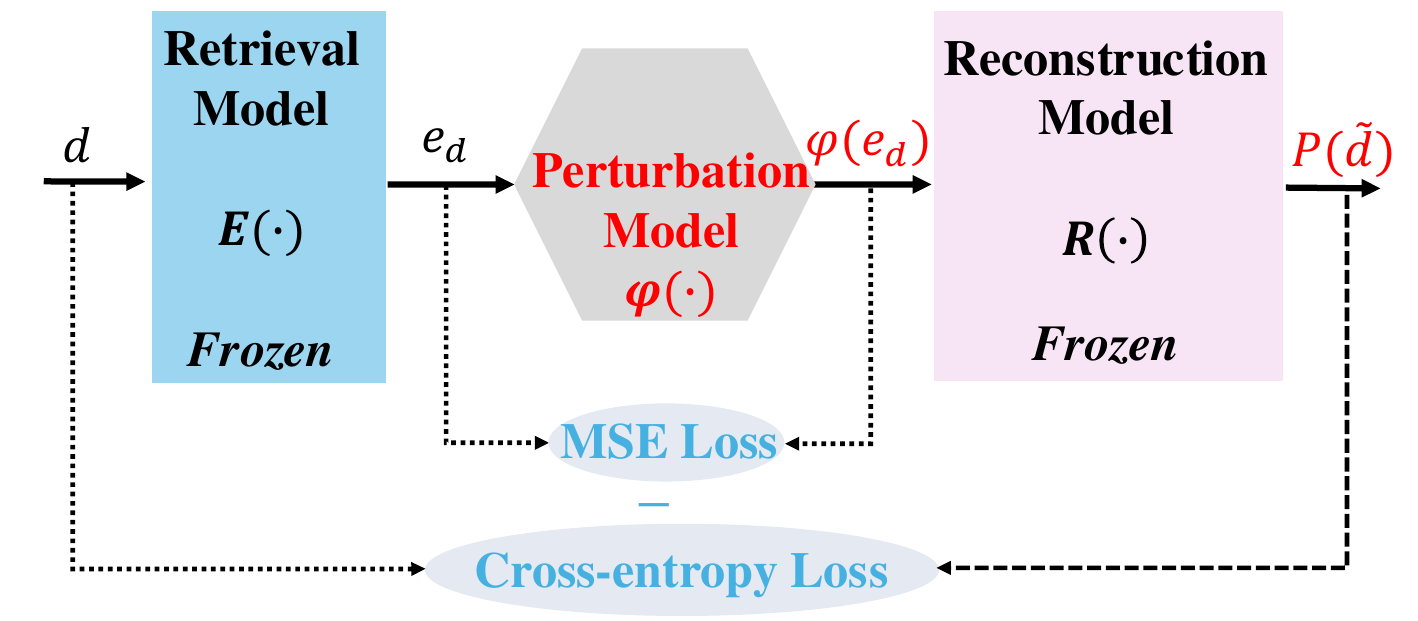}
  \caption{The pipeline of generating new adversarial documents.
  The perturbation model is trained using a combined loss to transform the embeddings of target documents into adversarial ones. Then, the trained reconstruction model recovers adversarial embeddings to adversarial documents.}
  \label{fig:IEM}
\end{figure}

To achieve these two conditions simultaneously, we use a perturbation mode~\(\varphi(\cdot) \) and design an adversarial generation optimization process with two loss functions, which is shown in Figure~\ref{fig:IEM}.
In this process, the input is a target document~\(d\), and the output is its adversarial document~\(\tilde{d}\).


To optimize the perturbation mode~\(\varphi(\cdot) \) for the  embedding similarity condition, we \textbf{minimize} the \underline{m}ean \underline{s}quare \underline{e}rror (\textbf{MSE}) loss between two embeddings:
\begin{align}
    L_{MSE} = \mathbb{E}\left[  \left( \varphi\left(e_{d}\right) -e_{d}  \right)^2 \right]   
\end{align} 

To implement the semantic irrelevance condition and 
maximize the token-level dissimilarity between target document~\(d\) and its adversarial document~\(\tilde{d}\),
we \textbf{maximize} the {\underline{c}ross-\underline{e}ntropy}~(\textbf{CE}) loss at token level as follows:
\begin{align}
\begin{split}
    L_{CE} = d\cdot \log \left( P\left( \tilde{d} \mid e_{d} ; \varphi \right) \right)  
\end{split}
\end{align}
where \(P\left( \tilde{d} \mid e_{d} ; \varphi \right) \in \mathbb{R}^{\left|d\right|\times\left|V\right|}\) is predicted probabilities of each token in the adversarial document~\(\tilde{d}
\), abbreviated as \( P\left( \tilde{d} \right) \). 
The perturbation model \(\varphi(\cdot) \) is optimized to \textbf{minimize} the  following loss function:
\begin{align}
    L_{Attack} = L_{MSE} - L_{CE} 
\end{align}
due to the differing scales of these two losses~( $L_{CE} \in [0,\infty) $), we use a hyperparameter \(\lambda =5 \) by default to clip the cross-entropy loss in our implementation, which allows more effective optimization, as shown in the following:
\begin{align}
    L_{Attack} = L_{MSE} - \min\left( L_{CE},  \lambda \right)
    \label{eq:attack_loss}
\end{align}
where \(\lambda\) controls the degree of semantic similarity at the token level. A larger \(\lambda\) value indicates a smaller token overlap between \(d\) and its adversarial document~\(\tilde{d}\).

In this paper, we use a three-layer perceptron to implement the perturbation model. Notably, although the perturbation model described above is document-specific, meaning a unique attack model is initialized for generating each adversarial document, its lightweight design with few parameters allows for fast and efficient training. 
Throughout the optimization process using the \(L_{Attack}\) loss, the model updates prediction probabilities~\(P\left(\tilde{d}\right)\) continuously and we can select tokens of the output adversarial document
by \(\tilde{d} = \argmax_{p(\tilde{t}_i ) \forall \tilde{t}_i \in V } P\left( \tilde{d}\right)\).

\begin{table*}[tbp] 
\caption{Statistics of datasets used in our work~\cite{thakur2021beir}. 
Avg. D/Q indicates the average number of relevant documents per query.} 
\small
\centering
\renewcommand\arraystretch{0.9} %
\setlength{\tabcolsep}{1mm}{ %
\resizebox{1.\linewidth}{!}{
\begin{tabular}{lcccccccccccccc}
\toprule
 \multirow{2}{*}{Datasets}  & \multirow{2}{*}{Task} & \multirow{2}{*}{Domain} & \multirow{2}{*}{Title}  & \multirow{2}{*}{Relevancy}  & \multirow{2}{*}{\#Corpus}    &\multicolumn{1}{c}{Train}  &\multicolumn{2}{c}{Test or Dev} &\multicolumn{2}{c}{Avg. Word Lengths} 
    \\\cmidrule(lr){7-7}\cmidrule(lr){8-9} \cmidrule(lr){10-11}
           &  &  &  & & &\#Pairs  &\#Query &Avg.D/Q   &Query  &Document \\
\hline
MS MARCO   &Passage-Retrieval & Misc.  & \texttimes &Binary &8,841,823 &532,761 &6,980  &1.1 &5.96&55.98 \\
TREC DL 19   &Passage-Retrieval & Misc.  & \texttimes &4-level &8,841,823 &532,761 &43  &95.4 &5.96&55.98 \\
TREC DL 20   &Passage-Retrieval & Misc.  & \texttimes &4-level &8,841,823 &532,761 &54  &66.8 &5.96 &55.98 \\
NQ   &Question Answering  & Wikipedia & \checkmark &Binary &2,681,468 &132,803 &3,452 &1.2 &9.16 &78.88   \\
Quora   & Retrieval &Quora  & \texttimes &Binary &522,931 &--- & 10,000&1.6 &9.53 &11.44 \\
FiQA & Question Answering &Finance  & \texttimes &Binary &57,638 &14,166 &648 &2.6 &10.77 &132.32 \\
Touché-2020 & Retrieval &Misc.  & \checkmark &3-level &382,545 & ---&49 &19.0 &6.55 &292.37 \\
\bottomrule
\end{tabular}}
}
 \label{tab:datasets_details}
\end{table*}

\section{Experiments}

In this section, we demonstrate the effectiveness and efficiency of the proposed method through extensive experiments conducted on two attack tasks and multiple datasets.

\subsection{Experimental Setup}

In this subsection, we outline the experimental setup, covering datasets, retrievers, evaluation metrics, and implementation details.\looseness=-1

\subsubsection{Datasets}
Since all retrievers in this paper are fine-tuned on the MS MARCO-Passage-Ranking dataset~(\textbf{MS MARCO})~\cite{bajaj2016ms}, we use its entire corpus to train the reconstruction model. And then the reconstruction models are tested on the corpus of Natural Questions(\textbf{NQ})~\cite{KwiatkowskiPRCP19_NQ}, another widely used dataset.

For adversarial attacks, we select \textbf{TREC DL 19}~\cite{Nick_trec_dl19} and \textbf{TREC DL 20}~\cite{CraswellMMYC20_trec_dl20} as in-domain target datasets as they share the same corpus with MS MARCO. Additionally, we select \textbf{NQ}, \textbf{Quora}~\cite{quora2017dataset}, \textbf{FiQA}~\cite{MaiaHFDMZB18_fiqa}, and \textbf{Touché-2020}~\cite{BondarenkoFBGAP20a_touche2020} from the \textbf{BEIR}~\cite{thakur2021beir} benchmark as out-of-domain datasets due to the diverse performance of retrieval models across these datasets. The statistics of these datasets are shown in Table~\ref{tab:datasets_details}.

\subsubsection{Retrieval Models}
We select \textbf{SimLM}\footnote{\url{https://huggingface.co/intfloat/simlm-base-msmarco-finetuned}}~\cite{WangYHJYJMW23_SimLM} as the primary target attack retriever as it represents one of the SOTA bi-encoder retrievers.
We also select \textbf{Contriever}~\cite{izacard2021contriever}, \textbf{E5-base-v2}~\cite{WangE5-base}, \textbf{TAS-B}~\cite{HofstatterLYLH21_tasb_dense_retrieval}, \textbf{DRAGON+}~\cite{LinALOLMY023_dragon}, and \textbf{RetroMAE}~\cite{XiaoLSC22_RetroMAE} as alternative attack targets and train their reconstruction models, respectively. All the retrievers aforementioned are fine-tuned on the MSMARCO dataset, distinguishing them from their respective pre-trained models.

For black-box attacks, we select four retrieval models with different structures as targets: \textbf{DPR}~\cite{KarpukhinOMLWEC20_DPR} (bi-encoder), \textbf{SimLM re-ranker} (cross-encoder), \textbf{ColBERTv2}~\cite{SanthanamKSPZ22_ColBERTv2} (late interaction), and \textbf{RankLLaMA}~\cite{MaWYWL24_RankLLaMA} (generative model).
All retrieval models used in this paper are publicly available and frozen without additional training.

\subsubsection{ Baseline Methods}

In this paper, we use three adversarial attack methods as baselines:

\textbf{Random Noise}:  Following~\cite{MorrisKSR23_vec2text,abs-2402-12784_Vec2Text_Understanding}, we add Gaussian noise on token-level embedding to replace our attack model.
\begin{equation}
     \varphi\left(e_{d_i}\right) = e_{d_i} +  \beta \cdot \epsilon, \; \epsilon  \sim \mathcal{N}(0, 1)
\end{equation}
where \(\beta\) is a hyperparameter controlling the injected noise amount. We select \(\beta=0.5\) by default following the search method in ~\cite{MorrisKSR23_vec2text}.

\textbf{Random Token}: We randomly replace tokens in the target document with arbitrary random tokens at a ratio of \( p =0.3 \), where the ratio is searched in a similar way to Random Noise.

\textbf{HotFlip-based}: It is one of the most widely used gradient-based word substitution methods and serves as a foundational technique for many other approaches in generating adversarial documents~\cite{ZhongHWC23_Poisoning,10.1145/3576923_PRADA,10.1145/3583780.3614793_MCARA,su2024corpus,Liu0GR0FC23_TARA}. In this paper, we refer to Zhong et al.~\cite{ZhongHWC23_Poisoning} and use the codebase from Li et al.~\cite{li2025reproducinghotflip}, as it is an accelerated version.

\subsubsection{Evaluate Metrics}

We use Normalized Discounted Cumulative Gain (specifically, \textbf{nDCG@10}) for retrieval performance. \looseness=-1

To evaluate the reconstructed model, we use four widely used metrics: \textbf{Accuracy}, \textbf{Precision}, \textbf{Recall}, and \textbf{F1} for this token-level multi-class classification task. It is worth noting that we report the macro-averaged scores for Precision, Recall, and F1.

To evaluate the attack performance, we assess from two perspectives: Embedding Similarity and Semantic Irrelevance, which align with the two conditions for attack success. 

\begin{figure}[t]
\centering
  \includegraphics[width=\columnwidth]{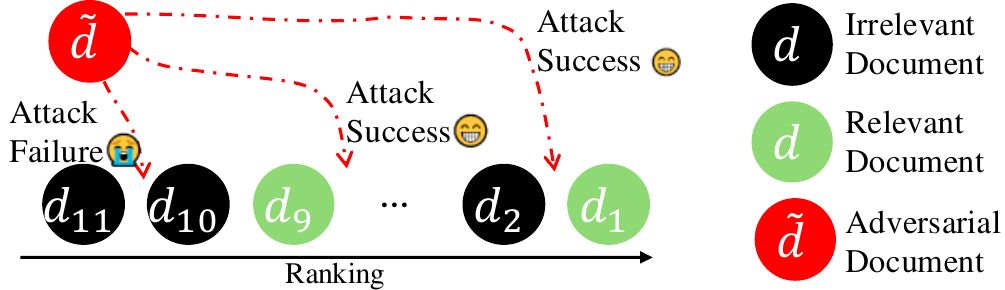}
  \caption{Definition of Attack Success Rate~(ASR). For an arbitrary ranking, if there is an adversarial document \(\tilde{d} \) that exceeds the relevant document with the lowest score \(d_9\), this ranking is considered as attacked successfully. }
  \label{fig:SR}
\end{figure}

For Embedding Similarity, our evaluation uses two metrics: Attack Success Rate~(\textbf{ASR}) and \textbf{Top@k}, which are defined as follows: 

$\bullet$ \textbf{ASR}: The attack success rate in this paper is proposed as the ratio of rankings for which the rank of relevant documents is affected by adversarial attacks, as illustrated in Figure~\ref{fig:SR}.
Its practical significance lies in representing the proportion of rankings in which the user encounters adversarial documents before obtaining all relevant documents.

$\bullet$ \textbf{Top@k}: It is one of the most commonly used in the literature~\cite{ZhongHWC23_Poisoning,li2025reproducinghotflip,anonymous2024gasliteing,10.1145/3548606.3560683_Order-Disorder,Liu0GRFC24_multi_granular}, which shows the ratio of queries that have at least one adversarial document in its top-$k$ retrieval result. We select Top@10 and Top@50 in this paper.

For Semantic Irrelevance, we use: the BiLingual Evaluation Understudy~(\textbf{BLEU}\footnote{\url{https://huggingface.co/spaces/evaluate-metric/bleu}})~\cite{PapineniRWZ02_BLEU} score and questions~(\textbf{Q1} and \textbf{Q2}), which are defined as follows:

$\bullet$ \textbf{BLEU}: it is a widely used word-level metric for evaluating the correspondence between a machine’s output and that of a human.

$\bullet$ \textbf{Questions}: we design two question prompts~(\textbf{Q1} and \textbf{Q2}) and then ask  Large Language Model~(LLM) and human expert to answer them.  The details of these two questions are shown in Table~\ref{table:llm prompts}. We report the ratio of ``\textbf{NO}'' response, the higher the better, as our adversarial documents would meet the semantic irrelevance condition. We select the latest GPT-4o-mini\footnote{\url{https://platform.openai.com/docs/models/gpt-4o-mini}}~\cite{abs-2303-08774_GPT-4} for  LLM evalution.

In addition, we also use the \textbf{Perplexity}\footnote{\url{https://huggingface.co/spaces/evaluate-metric/perplexity}} metric, calculated by the LLaMA-3.2 1B model~\cite{Llama3_head}, to assess the fluency of the adversarial documents. A lower perplexity value indicates higher fluency in
the adversarial document. \looseness=-1

\subsubsection{Implementation Details}
To train the reconstruction model, we use a learning rate of 1e-5, train for 5 epochs, and set the batch size to 128\(\times\)4 GPUs, with a maximum text length of 128. 
For all attack experiments, we set \(\lambda = 5 \), the learning rate is 5e-4. We run both our method and the HotFlip-based methods for 3000 epochs.  The HotFlip-based adversarial document is initialized as a $\texttt{[MASK]}$ token list of the same length as the target document.
Our experiments are mainly implemented using Pytorch 2.1 on a Ubuntu server with NVIDIA L4 ×24G × 8 GPUs, AMD EPYC 9554P CPU, and 384G memory.
Our code is available at \textbf{\url{https://github.com/liyongkang123/unsupervised_corpus_poisoning}}.

\begin{table}[tbp]
\caption{Two simplified question prompts used for LLM and Human semantic evaluation. We report the ratio of ``NO'' in their responses.  }
\renewcommand\arraystretch{1} %
\setlength{\tabcolsep}{2mm}{ %
\begin{tabular}{lp{6.5cm}}
\toprule
\multicolumn{1}{c}{{Question}} & \multicolumn{1}{c}{{Prompt}} \\
\hline
\multirow{5}{*}{\makecell{Q1}}  &You are an expert in relevance assessment. I will provide
you with a query posed by a user, followed by a document.
Your task is to determine \textbf{whether the document answers the
user's question}. Please respond directly and solely with ``Yes''
or ``No''.   \\
\hline
\multirow{5}{*}{\makecell{Q2}} &You are an expert in relevance assessment. I will provide
you with two documents, and you need to assess \textbf{whether
these two documents express the same information}. Please
respond directly and only with ``Yes'' or ``No''. \\
\bottomrule
\end{tabular}}
\label{table:llm prompts}
\end{table}

\begin{table}[t]
\caption{Performance of reconstruction models tested on the NQ corpus. 
We report the macro-averaged score for the Precision, Recall, and F1.   All retrievers come from fine-tuned versions on the MSMARCO  rather than raw pre-trained models. \looseness=-1
}
\renewcommand\arraystretch{1} %
\setlength{\tabcolsep}{2mm}{ %
\begin{tabular}{lccccccc}
\toprule
Retrievers   & Accuracy\(\,\uparrow\)  & Precision\(\,\uparrow\) & Recall\(\,\uparrow\)  &F1\(\,\uparrow\)  \\
\hline
SimLM &0.9914  &0.9449 &0.9501 &0.9463  \\
Contriever  &0.9880  &0.9425 &0.9458   &0.9430 \\
E5-base-v2  &0.9778  &0.9245  &0.9098   &0.9135 \\
TAS-B  &0.9915   &0.9443   &0.9522   &0.9471 \\ 
DRAGON+   & 0.9899 &0.9416   &0.9414   &0.9399 \\
RetroMAE &0.9911 &0.9373 &0.9426 &0.9386 \\
\bottomrule
\end{tabular}}
\label{out of domain_RM}
\end{table}

\begin{table*}[t] 
\caption{The Top-1 white-box attack performance for attacking SimLM on six datasets.
We report not only on two attack success conditions—embedding similarity and semantic irrelevance—but also on the perplexity and time cost for generating each adversarial document (measured in seconds). All results are averaged over three runs with different random seeds.
} 
\centering
\renewcommand\arraystretch{0.86} %
\setlength{\tabcolsep}{1.1mm}{ %
\begin{tabular}{l|cccccccrr}
\toprule
 \multirow{2}{*}{Datasets} & \multirow{2}{*}{\makecell[cc]{Attack \\ Methods}}   & \multicolumn{3}{c}{Embedding Similarly }     & \multicolumn{3}{c}{Sematic Irrelevance} & \multirow{2}{*}{Perplexity\(\,\downarrow\)}& \multirow{2}{*}{{Time Cost (s)}\(\,\downarrow\)}
    \\\cmidrule(lr){3-5} \cmidrule(lr){6-8} 
                &  &ASR\(\,\uparrow\) & Top@10\(\,\uparrow\)& Top@50\(\,\uparrow\) &BLEU\(\,\downarrow\)  & LLM Q1\(\,\uparrow\) &LLM Q2\(\,\uparrow\)   \\            
\hline
\multirow{4}{*}{\makecell[cc]{\textbf{TREC DL 19 }\\ nDCG@10$=$0.650}} & Random Noise  &0.735 &0.023  &0.046 &0.037 &\textbf{1.000}  &\textbf{0.984} &3469.2 & 0.1\\
&Random Token  &0.736 &0.147  &0.357 &0.349 &0.930  &0.566 &1638.2  & \textbf{0.1} \\
&HotFlip-based &0.899 &\textbf{0.682}  &0.767 &\textbf{0.020} & 0.992 &0.861 &6032.6 & 512.1 \\
&Ours  &\textbf{0.984} &0.434  &\textbf{0.791} &0.093 &0.954  &0.752 &\textbf{188.9} &119.9 \\ 
\hdashline
\multirow{4}{*}{\makecell[cc]{\textbf{TREC DL 20} \\ nDCG@10$=$0.639} } & Random Noise &0.679 &0.012  &0.062 &0.027 &\textbf{1.000}  &\textbf{0.982}  &3410.3   & 0.1\\
&Random Token  &0.710 &0.185  &0.383 &0.355 &0.907  &0.642 &1793.4   & \textbf{0.1}   \\
&HotFlip-based   &0.778 &\textbf{0.617}  &0.673 &\textbf{0.019} &0.961  &0.747 &6390.9 &449.8 \\
&Ours  &\textbf{0.957} &0.500  &\textbf{0.753} &0.079 &0.944  &0.753 & \textbf{166.0} & 115.6 \\
\hdashline
\multirow{4}{*}{\makecell[cc]{\textbf{NQ }\\ nDCG@10$=$0.426}}& Random Noise   &0.063 &0.030  &0.100 &0.052 &\textbf{1.000}  &\textbf{0.993}  &3001.5  &\textbf{0.1 }\\
&Random Token  &0.177 &0.217  &0.4 &0.348 &0.940  &0.597   &2889.4   &0.1 \\
&HotFlip-based   &\textbf{0.557} &\textbf{0.773}  &\textbf{0.817} &\textbf{0.025} &0.983  &0.830 &8021.7  &588.3 \\
&Ours  &0.417 &0.490  &0.700 &0.099 & 0.940 &0.793 &\textbf{231.2} &127.5 \\
\hdashline
\multirow{4}{*}{\makecell[cc]{\textbf{Quora} \\ nDCG@10$=$0.865}} & Random Noise  &0.003 &0.027  &0.047 &\textbf{0.040} & \textbf{1.000} &\textbf{1.000 }&4104.9  & \textbf{0.1} \\
&Random Token   &0.057 &0.187  &0.287 &0.349 &0.967  &0.883 &10265.8  &0.1   \\
&HotFlip-based   &\textbf{0.140} &\textbf{0.527}  &\textbf{0.593} &0.074 &0.970  &0.730 &59468.5  & 136.5 \\
&Ours   &0.043 &0.320  &0.513 &0.051 & 0.993 &0.923 &\textbf{461.7} &96.5 \\
\hdashline
\multirow{4}{*}{\makecell[cc]{\textbf{FiQA} \\ nDCG@10$=$0.224} } & Random Noise  &0.483 &0.047 &0.173 &0.074   &\textbf{1.000} &\textbf{0.980} &2532.8  &\textbf{0.1}\\
&Random Token  &0.387 &0.103 &0.220 &0.335  &0.997 &0.830 &1664.4  &0.1 \\
&HotFlip-based  &0.660 &\textbf{0.580}  &0.640 &\textbf{0.013} &1.000  &0.960 &6072.2 &770.3 \\
&Ours   &\textbf{0.760} &0.370  &\textbf{0.643} &0.102 &0.983  &0.833 &\textbf{159.0} &138.7 \\
\hdashline
\multirow{4}{*}{\makecell[cc]{\textbf{Touché-2020} \\ nDCG@10$=$0.162} } & Random Noise  &0.408 &0.000 &0.020 &\textbf{0.038}   &\textbf{1.000} &\textbf{0.986} &4495.5  &0.1  \\
&Random Token  &0.639 &0.101 &0.245 &0.353  &0.939 &0.748 &7366.4 &\textbf{0.1} \\
&HotFlip-based  &0.666 &\textbf{0.449}  &0.544 &0.051 &1.000  &0.735 &14037.5  &231.0    \\
&Ours   &\textbf{0.986} &0.422  &\textbf{0.633} &0.086 &0.959  &0.803 &\textbf{522.2} &102.4 \\
\bottomrule
\end{tabular}}
 \label{tab: top1_attack_simlm_attack_white}
\end{table*}

\subsection{Reconstruction Model Performance}

For each retriever \(E(\cdot) \), we fine-tune a corresponding reconstruction model \(R(\cdot)\) on the MS MARCO corpus, initializing it from an uncased BERT large model. 
Table~\ref{out of domain_RM} shows the reconstruction capability of our method, tested on the NQ corpus (out-of-domain). 

It can be observed that reconstruction models 
effectively learn the relationship between contextual token-level embeddings and texts, with an Accuracy of around 0.99. While the overall F1 score appears relatively low, it still averages above 0.93. This is due to the differences between the NQ corpus and MS MARCO, which pose challenges for the model in accurately predicting tokens with extremely low frequency. Consequently, this discrepancy results in a lower macro-averaged score. 

\subsection{Top-1 Attack}\label{sec:top-1 attack}

In this section, we introduce the Top-1 attack, which simulates real-world scenarios by attacking a ranking.
We select SimLM, as the target model to attack.
We randomly sampled up to 100 test queries from each of the 6 datasets, using SimLM to retrieve documents. The top-ranked document \(d\) was input into our perturbation method and three baseline methods, which generated the adversarial document~\(\tilde{d}\)  and inserted into the corpus. This adversarial document~\(\tilde{d}\) is expected to be retrieved at a very high rank in that ranking. It must be noted that the top-1 attack here targets a ranking, which in real-world scenarios does not necessarily originate directly from a specific query or multiple queries. 

\begin{figure*}[t]
    \centering
  \includegraphics[width=2\columnwidth]{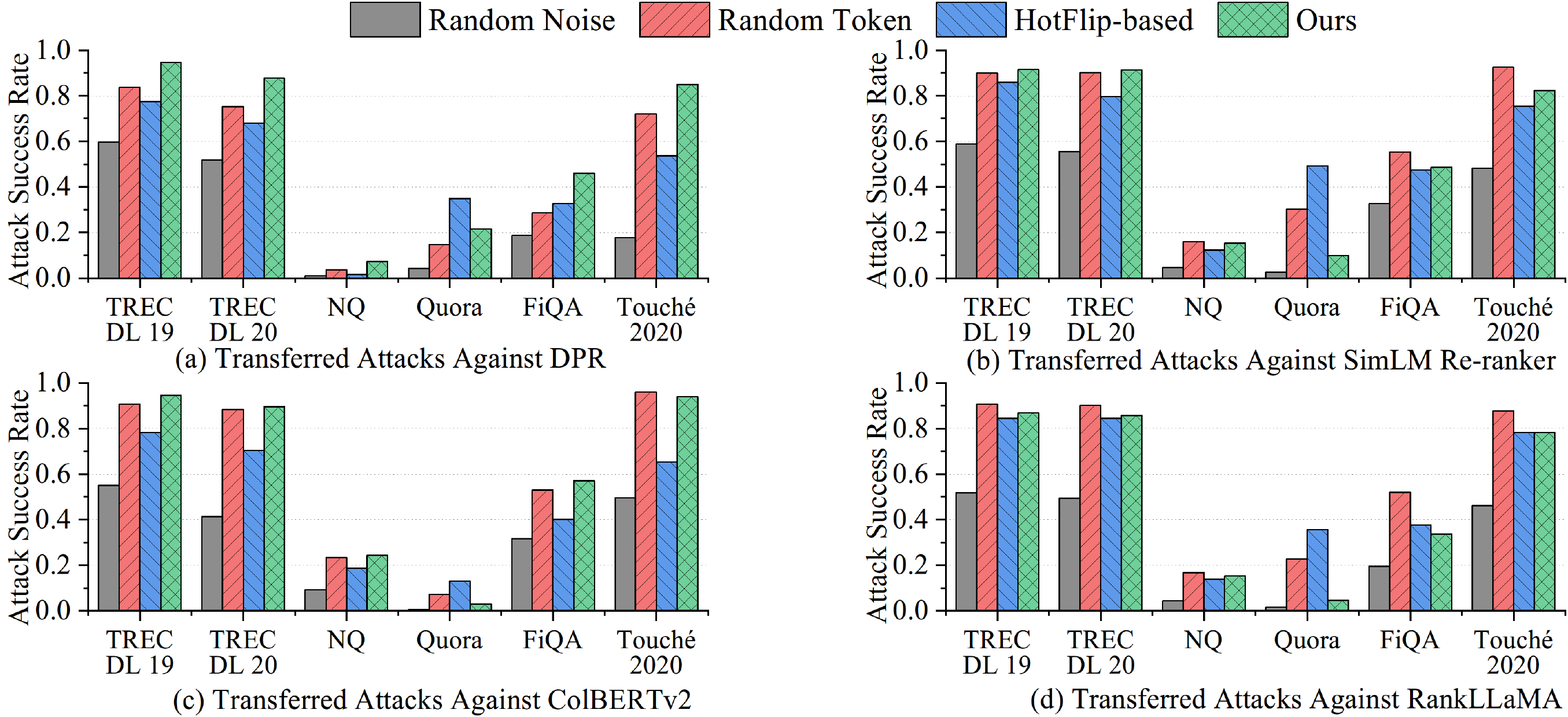}
  \caption{The Top-1 black-box attack results by transferring adversarial documents from SimLM to other retrieval models.}
  \label{fig:attack_transfer_simlm_to_models}
\end{figure*}

\subsubsection{White-box Attack Performance}

We first conduct experiments in a white-box setting, where the target retriever is known—in our case, SimLM.
The results of the attack are presented in Table~\ref{tab: top1_attack_simlm_attack_white}, where we have the following observations:

$\bullet$ Under the three metrics of Embedding Similarity—ASR, Top@10, and Top@50—Ours and the HotFlip-based method perform comparably across the six datasets. The Random Noise and Random Token methods exhibit the poorest performance due to their inherent randomness and lack of control.

$\bullet$ Under the Semantic Irrelevance metric, Random Noise performs the best because adding small-scale random noise makes the embedding represent a document entirely unrelated to the target document, maximizing semantic irrelevance.  We can also observe that the Random Token method performs poorly because, after replacing 30\% of the tokens, 70\% remain unchanged, resulting in a relatively high BLEU score (around 0.35).
Moreover, we observe that HotFlip achieves lower LLM Q2 scores when attacking Quora. Upon closer inspection, this is due to HotFlip's tendency to select words consistent with the original document during gradient computation when the target document is short.
Overall, Ours and the HotFlip-based method still perform comparably for semantic irrelevance. \looseness=-1

$\bullet$ For the perplexity metric, our method achieves significantly lower scores than the other three methods, indicating that our adversarial documents are more difficult to detect using perplexity-based filters.
Notably, even with only 30\% of tokens altered, the Random Token method produces remarkably high perplexity scores on the LLama-3.2 1B model. 
The low perplexity of our method is likely attributed to the reconstruction model, where each token is predicted based on its contextual embedding derived from the entire adversarial document during reconstruction.

$\bullet$ For the time cost, Random Noise, and Random Token require negligible processing time, while HotFlip is four times slower than our method on average. Notably, we use the fastest accelerated  HotFlip implementation ~\cite{li2025reproducinghotflip} here, whereas Zhong et al.~\cite{ZhongHWC23_Poisoning} require over 2 hours per document~\cite{abs-2402-12784_Vec2Text_Understanding}.

Considering all the results in Table~\ref{tab: top1_attack_simlm_attack_white}, our method matches the SOTA HotFlip-based model in attack effectiveness, while producing adversarial documents with significantly lower perplexity and four times higher efficiency.

\subsubsection{Black-box Attack Performance}

To evaluate the effectiveness of our method in a transfer-based black-box attack setting, where the target retrieval model is unknown, we use the same sampled queries and their generated adversarial documents from the previous white-box attack section. Then we test the attack success rate based on the retrieval ranking from four retrieval models: DPR, SimLM re-ranker, ColBERTv2, and RankLLaMA. 

It is worth noting that, in the literature~\cite{10.1145/3576923_PRADA,Liu0GR0FC23_TARA,Liu0GRFC24_multi_granular,10.1145/3583780.3614793_MCARA}, black-box attacks typically involve distilling a surrogate model first, then using white-box methods to attack that surrogate model.
However, distilling LLMs like LLaMA is too computationally intensive in our experiments. Therefore, to ensure fairness across all target models and to reduce the computational workload, we refrain from training surrogate models. Instead, we follow the approach in ~\cite{li2025reproducinghotflip} and directly use the adversarial documents generated by various attack methods during attacks on SimLM and apply them to the target models. \looseness=-1

The results, averaged over three runs, are shown in Figure~\ref{fig:attack_transfer_simlm_to_models}.
We can observe that, when transferred to four black-box models, the attack success rate of our method surpasses that of the HotFlip-based method on most datasets, indicating better transferability. We speculate that this is because our adversarial documents have lower perplexity~(and potentially higher fluency), comparable to normal texts, making them more effective at deceiving other retrieval models.\looseness=-1

We also find that the Random Token method has the highest attack success rate. However, as shown in Table~\ref{tab: top1_attack_simlm_attack_white}, the BLEU scores of the adversarial documents generated by Random Token are very high, which does not satisfy the condition of semantic irrelevance. Therefore, considering both the ASR and BLEU metrics, the Random Token attack is not an effective method.

Another interesting finding is that, when we compare the four target models, we observe that the success rate of all attack methods transferred to the DPR model is almost always lower than that of SimLM re-ranker and ColBERTv2. This indicates that the robustness of the DPR model is higher than that of SimLM re-ranker and ColBERTv2. Similarly, RankLLaMA also demonstrates high robustness, as its success rate after being attacked is relatively low.

\begin{table}[tbp] 
\caption{Human and LLM evaluation on semantic irrelevance.
} 
\centering
\renewcommand\arraystretch{0.9} %
\setlength{\tabcolsep}{3mm}{ %
\begin{tabular}{l|ccccccccccccc}
\toprule
\multirow{2}{*}{\makecell[cc]{Attack \\ Methods}}   & \multicolumn{2}{c}{Question 1}     & \multicolumn{2}{c}{Question 2} 
    \\\cmidrule(lr){2-3} \cmidrule(lr){4-5} 
                &LLM & Human & LLM  &Human \\ 
\hline
Random Noise  &1.000 & 1.000  &0.967  &0.989 \\
Random Token  &1.000 &0.900  &0.767 &0.922  \\
HotFlip-based & 1.000 &0.867 & 0.800 &0.853\\
Ours  &1.000 &0.900  &0.830 & 0.930\\ 
\bottomrule
\end{tabular}}
 \label{tab: top1_attack_tas-b_attack_part_NEW}
\end{table}

\begin{figure*}[t]
\centering
  \includegraphics[width=2\columnwidth]{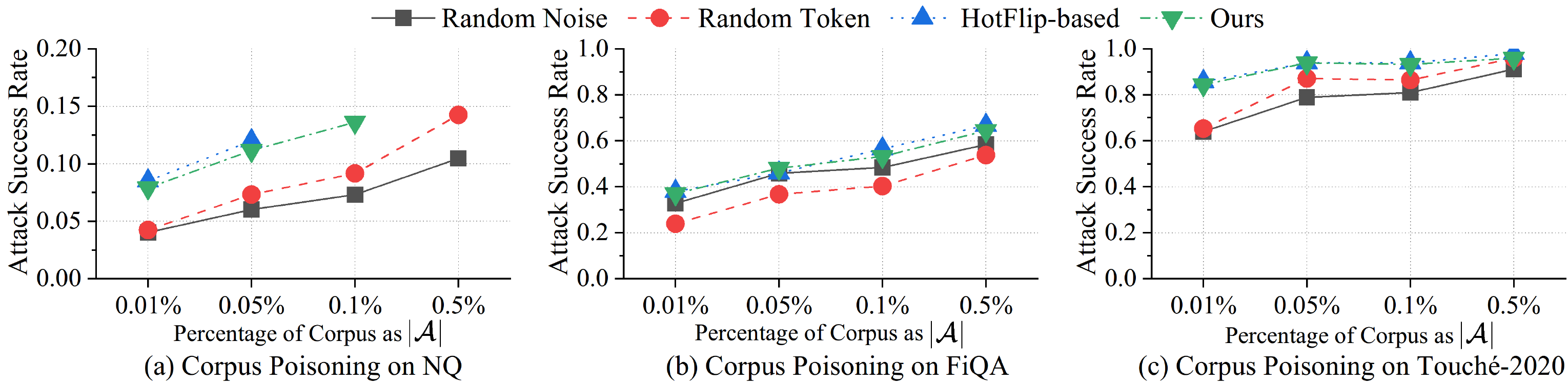}
  \caption{Corpus poisoning attack results on three datasets.  Some data points are not included in NQ due to computational complexity.  
    The number of injected adversarial documents $|\mathcal{A}|$ is determined by multiple percentages of corpus size.
  }
  \label{fig:corpus attack results}
\end{figure*}

\subsubsection{Human evaluation Vs. LLM evaluation}

In this paper, we primarily use LLMs to evaluate semantic irrelevance from a human perspective, as LLMs are not only much cheaper and easier to test on a large scale but are also highly effective at making judgments~\cite{ChiangL23,RahmaniYC00CASF24}.
 However, human evaluation is still necessary, as we are unclear whether humans and LLMs align consistently on this specific issue. Therefore, from all the documents generated by white-box attacks in Table~\ref{tab: top1_attack_simlm_attack_white}, we randomly select 30 adversarial documents from each attack method and invite three experts in the information retrieval field to evaluate Question 1 and Question 2. At the same time, we also use LLMs to evaluate these documents. 
 
 The experimental results are shown in Table 5. We can observe that for Question 1, LLMs exhibit stricter judgment, consistently answering ``No'' when uncertain, whereas humans are more lenient. For Question 2, LLMs are relatively more permissive, particularly when keywords are highly repetitive, often interpreting them as semantically similar (resulting in the lowest score for Random Token). Overall, Random Token and HotFlip perform poorly on both questions, while our method effectively preserves semantic irrelevance.\looseness=-1

\subsection{Corpus Poisoning Attack}

The Top-1 attack validated the effectiveness of our method when attacking an arbitrary ranking. However, there are situations where no target ranking exists. For these cases, we refer to it as a corpus poisoning attack task. Additionally, unlike the definition of
corpus poisoning in Zhong et al.~\cite{ZhongHWC23_Poisoning}, we follow the approach of Li et al.~\cite{li2025reproducinghotflip}, which requires that attacks be conducted without prior knowledge of the queries. This is because, in real-world scenarios, an attacker can obtain a sample of the corpus distribution through manual inputs and by observing the ranking results, while acquiring real queries presents greater challenges.

In this experiment, we attack SimLM on NQ, FiQA, and Touché-2020 due to their varying corpus sizes and apply attacks using all four attack methods.
We do k-means clustering on the corpus, clustering them into \(\left|\mathcal{A}\right|\) categories. All documents in each cluster can be considered a ranking, sorted by their distance to the cluster centroid. We attack the document closest to the centroid in each cluster~(top-1 document of each ranking), and obtain adversarial documents~\(\mathcal{A}=\left\{ \tilde{d_1},\dots , \tilde{d}_{\left|\mathcal{A}\right|}\right\}\). We use all test queries of these two datasets to evaluate the performances.
Due to the varying sizes of different corpus, we used the proportion of the corpus as the number of clustering clusters \(\left|\mathcal{A}\right|\) to ensure fairness.

The corpus poisoning attack results are shown in Figure~\ref{fig:corpus attack results}, where we have the following findings:

$\bullet$ 
The attack success rate generally increases with the number of clusters. Moreover, the performance of Ours and HotFlip-based methods is comparable, both outperforming the Random Noise and Random Token methods. This trend is consistent with the white-box Top-1 attack results in Table~\ref{tab: top1_attack_simlm_attack_white}.

$\bullet$  All methods show a higher attack success rate on Touché-2020 compared to their performance on FiQA and NQ. We speculate that this may be because Touché-2020 has a higher average number of relevant documents per query, making it more susceptible to attacks. 
Specifically, in Touché-2020, even with only 0.01\% of adversarial documents, a high attack success rate can be achieved.

 $\bullet$ On the NQ dataset, our method significantly outperforms random noise, which might indicate that as the corpus size increases, the random noise approach becomes less effective.


In summary, our method and the HotFlip-based method are comparable in terms of attack effectiveness. However, considering attack efficiency and perplexity, the method proposed in this paper still holds an advantage.

\section{Additional Discussion}

In this section, we conduct additional analysis to provide a comprehensive evaluation of the performance of our approach.

\subsection{Hyper-parameter Study}

In Equation~\ref{eq:attack_loss}, the loss function \(L_{Attack}\) has two components:
minimizing \(L_{MSE}\) ensures that the output embedding has the minimum Euclidean distance to the input, while maximizing \(L_{CE}\) encourages that the output embedding results in different tokens after reconstruction.
These two losses compete during training, making 
optimization challenging without effective regulation. 
We tested different weights for the losses and multi-task learning methods (e.g., MGDA~\cite{SenerK18_MGDA}), but the results were unsatisfactory. Ultimately, we found that truncating \(L_{CE}\) stabilized the model's output.\looseness=-1

\begin{figure}[tbp]
\centering
  \includegraphics[width=\columnwidth]{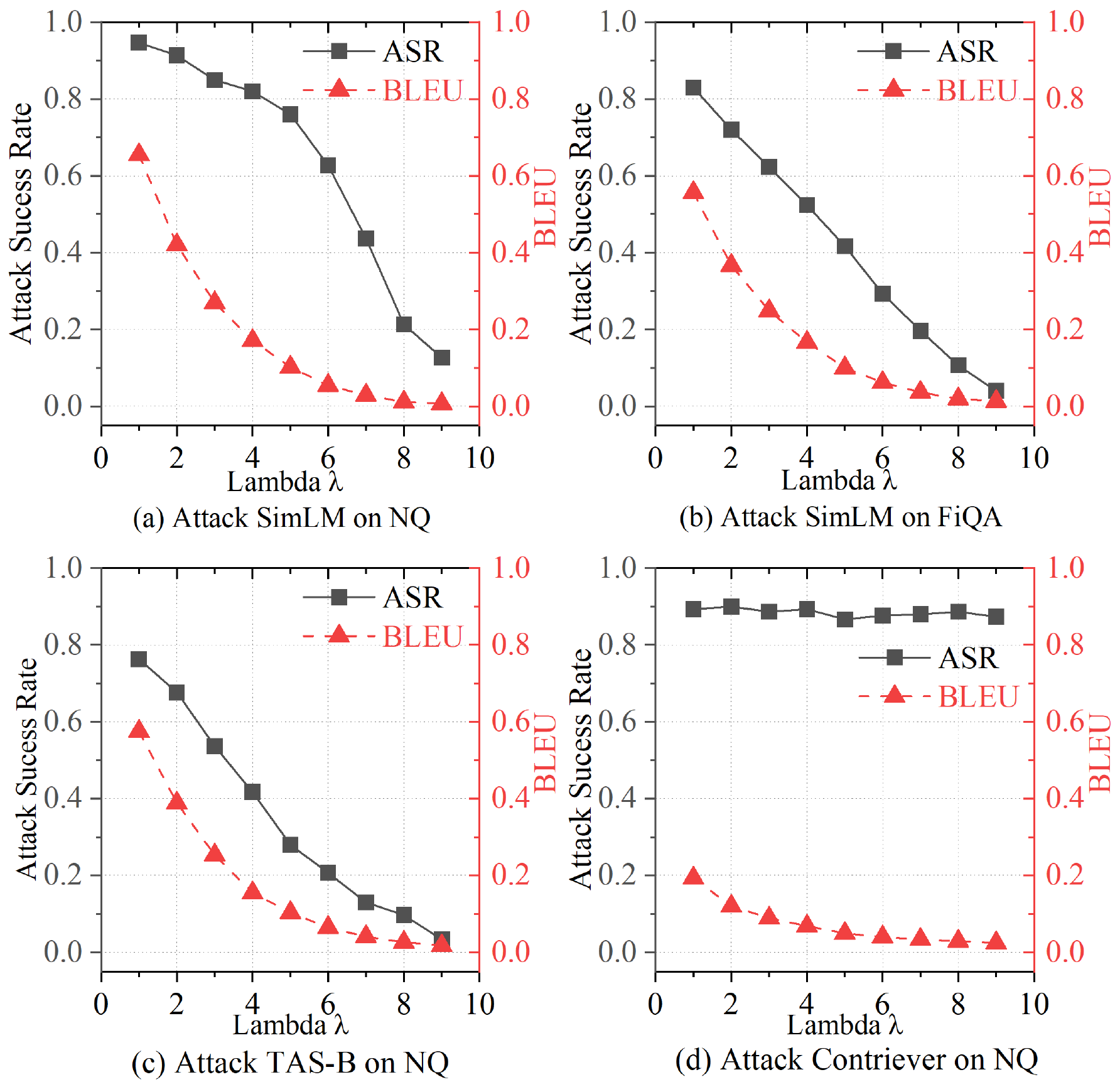}
  \caption{Hyper-parameter study about the trade-off of \(\lambda\).}
   \label{fig: lambda Trade-off}
\end{figure}

To demonstrate the effect of lambda \(\lambda\) on different datasets and retrievers, we selected the NQ and FiQA datasets and conducted experiments using lambda \(\lambda\) values from 1 to 9 on SimLM, TAS-B, and Contriever. The experimental results are shown in Figure~\ref{fig: lambda Trade-off}. 

By comparing all the subfigures in Figure~\ref{fig: lambda Trade-off}, we can observe that the BLEU score decreases monotonically with the increase of lambda, indicating that we can adjust the degree of semantic dissimilarity by controlling lambda. By comparing Figure~\ref{fig: lambda Trade-off} (a) and (b), we can observe that for attacking the same retriever on different datasets, \(\lambda=5\) is a suitable value to achieve both a high attack success rate and a relatively low BLEU score. By comparing Figure~\ref{fig: lambda Trade-off}(a), (c), and (d), we observe that \(\lambda=5\) exhibits strong robustness across attacks on various retrievers, highlighting its high generalizability.
Furthermore, in Figure~\ref{fig: lambda Trade-off}(d), we observe that the attack on Contriever always achieves a very high success rate, indicating that the Contriever model is highly vulnerable and susceptible to attacks. This conclusion aligns with the experimental findings in~\cite{ZhongHWC23_Poisoning,li2025reproducinghotflip}.

\subsection{Adversarial Training}

Adversarial training is a vital way to improve the robustness of retrieval models~\cite{10.1145/3477495.3531818,LupartC23_FGSM,ZhangGS0DC22,ParkC19}.
However, generating a large number of high-quality adversarial samples or hard negative samples has always been a challenging task. Given the efficiency of our method, we can create a substantial amount of adversarial samples offline for training, thereby enhancing the robustness of the model.
We offer a preliminary experiment: We select 7000 positive query-document pairs from the MS MARCO training set and generate one adversarial document for each positive document with our method. We then use the adversarially generated documents as negative samples, and fine-tune SimLM using a standard DPR training setting with a contrastive loss and in-batch negative samples~\citep{KarpukhinOMLWEC20_DPR}. 
We observe that the attack success rate of the Top-1 attack decreases across all datasets after adversarial training, with an average relative reduction of 7.9\%. This finding suggests that the model's resilience against our attacks has been enhanced, leading to improved robustness. Additionally, the fine-tuned SimLM model achieves a minor 0.002 increase in the average nDCG@10 across six datasets.

\begin{figure}[tbp]
\centering
  \includegraphics[width=\columnwidth]{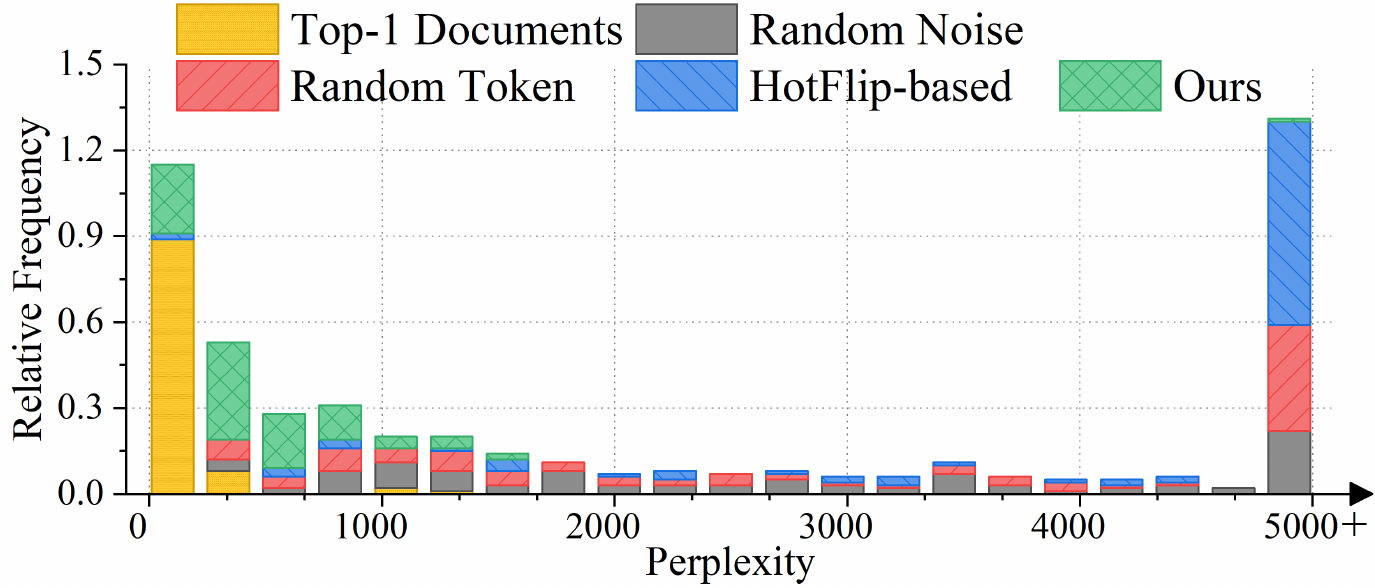}
  \caption{Stacked histogram showing the perplexity distribution of adversarial documents across all methods, with a maximum perplexity capped at 5000 for visualization clarity. \looseness=-1}
  \label{fig:Perplexity}
\end{figure}

\subsection{Perplexity Detection \& Case Study}
Figure~\ref{fig:Perplexity} illustrates the perplexity of the top-1 documents and adversarial documents generated by four different methods during the attack on Quora, as described in Section~\ref{sec:top-1 attack}. It can be observed that perplexity-based filtering~\cite{song2020adversarial} struggles to differentiate our method from the top-1 documents, whereas it easily distinguishes other methods, such as HotFlip. \looseness=-1

\begin{figure}[tbp]
\centering
  \includegraphics[width=\columnwidth]{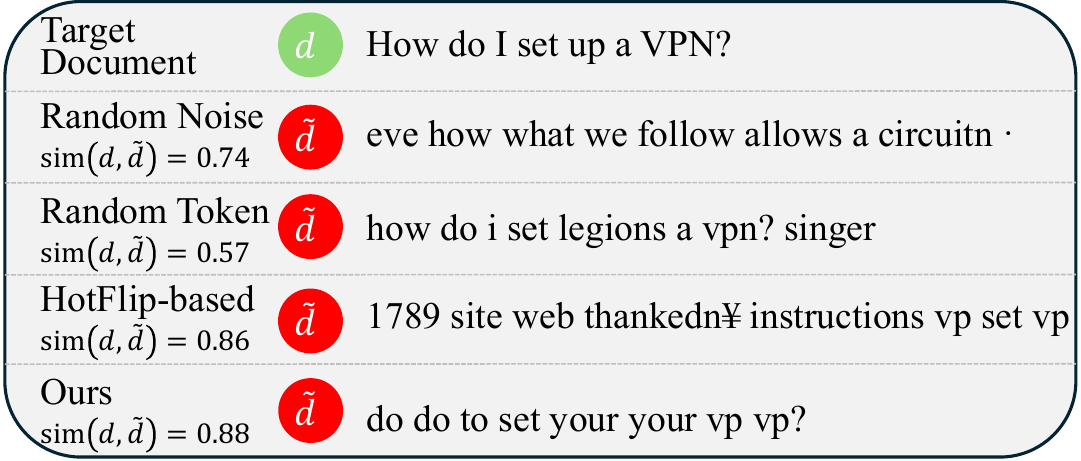}
  \caption{Case study from the Quora dataset (retrieval of similar questions). Target document ID: 182134.}
  \label{fig:case study}
\end{figure}

Figure~\ref{fig:case study} presents a case study of attacking the Quora dataset. 
It can be observed that although the documents generated by HotFlip-based methods exhibit high similarity, they are extremely disorganized, resulting in a high perplexity. In contrast, the random token method significantly reduces similarity after replacing a few tokens. However, our method achieves the best overall performance.

\section{Conclusions}
In this paper, we propose an unsupervised corpus poisoning task under a realistic attack setting. Adversarial documents are generated using our reconstruction and perturbation models, trained with the dual objective of maximizing the token-level dissimilarity while maintaining high embedding similarity. Our attack is fast, transferable, and shows that SOTA dense retrieval models are vulnerable. Our experiments include two scenarios: one where the top-1 document in a ranking is targeted, and another that targets a small percentage of a corpus, based on clustering.
Furthermore, our attack is hard to detect with perplexity metrics, as the adversarial examples generated, although nonsensical, follow the distribution of natural text more so than previous methods.
By leveraging the efficiency of our method, we enable adversarial training by generating adversarial documents on a larger scale, with preliminary results showing reduced attack success without harming retrieval performance.\looseness=-1

\begin{acks}
This work was partially supported by the China Scholarship Council (202308440220), the LESSEN project (NWA.1389.20.183) of the research program NWA ORC 2020/21 which is financed by the Dutch Research Council (NWO), and the PACINO project~(215742) which is financed by the Swiss National Science Foundation (SNSF).
\end{acks}

\bibliographystyle{ACM-Reference-Format}
\balance
\bibliography{sigir25}



\end{document}